# Communication Systems for Grid Integration of Renewable Energy Resources

F. Richard Yu[*], Peng Zhang[§], Weidong Xiao[#], and Paul Choudhury[+]

[*]Department of Systems and Computer Engineering
Carleton University, Ottawa, ON, Canada, K1S 5B6, Email: Richard_Yu@Carleton.ca

[§]Department of Electrical and Computer Engineering
University of Connecticut, Storrs, Connecticut 06269-2157, USA
Email: peng@engr.uconn.edu

[#]Program of Electrical Power Engineering, Masdar Institute of Science and Technology, Abu Dhabi, UAE
Email: mwxiao@masdar.ac.ae

[+]Asset Risk & Performance Management Department
BC Hydro, Vancouver, BC, Canada, V7X 1V5
Email: Paul.Choudhury@bchydro.com

*Abstract* — There is growing interest in renewable energy around the world. Since most renewable sources are intermittent in nature, it is a challenging task to integrate renewable energy resources into the power grid infrastructure. In this grid integration, communication systems are crucial technologies, which enable the accommodation of distributed renewable energy generation and play extremely important role in monitoring, operating, and protecting both renewable energy generators and power systems. In this paper, we review some communication technologies available for grid integration of renewable energy resources. Then, we present the communication systems used in a real renewable energy project, Bear Mountain Wind Farm (BMW) in British Columbia, Canada. In addition, we present the communication systems used in Photovoltaic Power Systems (PPS). Finally, we outline some research challenges and possible solutions about the communication systems for grid integration of renewable energy resources.

I.　INTRODUCTION

As concerns about climate change, rising fossil fuel prices and energy security increase, there is growing interest around the world in renewable energy resources. Since most renewable energy sources are intermittent in nature, it is a challenging task to integrate a significant portion of renewable energy resources into the power grid infrastructure. Traditional electricity grid was designed to transmit and distribute electricity generated by large conventional power plants. The electricity flow mainly takes place in one direction from the centralized plants to consumers. In contrast to large power plants, renewable energy plants have less capacity, and are installed in a more distributed manner at different locations. The integration of distributed renewable energy generators has great impacts on the operation of the grid and calls for new grid infrastructure. Indeed, it is a main driver to develop the *smart grid* for infrastructure modernization [1], which monitors, protects, and optimizes the operation of its interconnected elements from end to end with a two-way flow of electricity and information to create an automated and distributed energy delivery network.

Communication systems are crucial technologies for grid integration of renewable energy resources. Two-way communications are the fundamental infrastructure that enables the accommodation of distributed energy generation and assists in the reconfiguration of network topology for more efficient power flow. Many equipment in the grid (e.g., meters, sensors, voltage detection, etc.) should be monitored and controlled, which will enable important decision support systems and applications, such as Supervisory Control and Data Acquisition (SCADA), Energy Management System (EMS), protective relaying for high voltage lines, mobile fleet voice and data dispatch, distribution feeder automation, generating plant automation, physical security, etc. These applications are vital in monitoring, operating, and protecting both renewable energy generators and power systems.

There are several communication options available for the grid integration of renewable energy resources. These options include a hybrid mix of technologies, such as fiber optics, copper-wire line, power line communications, and a variety of wireless technologies. There is currently on-going debate surrounding what will emerge as the communications standard of choice. Since utilities want to run as many applications as possible over their networks, issues of bandwidth, latency, reliability, security, scalability and cost will continue to dominate the conversation. In addition, distinct characteristics in electric grid pose new challenges to the communication systems for grid integration of renewable energy resources.

In this paper, we will review some communication technologies available for grid integration of renewable energy resources. Then, we introduce a real renewable energy project, Bear Mountain Wind farm (BMW) in British Columbia, Canada, with thirty four ENERCON wind turbine generators and a generation capacity of 102 MW. Particularly, we present the communication systems used in BMW. In addition, we describe the communication systems used in Photovoltaic Power Systems (PPS). For both wind and photovoltaic power systems, we outline some research challenges and possible solutions about the communication systems for grid integration of these renewable energy resources.

The rest of this paper is organized as follows: Section II describes an overview of communication systems for grid integration of renewable energy resources. Section III presents the communication systems and some research challenges for grid integration of wind farms. Section IV presents the communication systems and some research challenges for grid integration of photovoltaic power systems. Finally, we conclude the paper in Section V.

II. OVERVIEW OF COMMUNICATION SYSTEMS FOR GRID INTEGRATION OF RENEWABLE ENERGY RESOURCES

A typical electric grid communication system consists of a high-bandwidth backbone and lower-bandwidth access networks, connecting individual facilities to the backbone. Fiber optics and/or digital microwave radio are usually the technologies for backbone, whereas the access may use alternatives such as copper twisted-pair wire lines, power line communications, and wireless systems. In this section, we introduce some communication technologies and related standards that are particularly interesting for grid integration of renewable energy resources.

*A. Power Line Communications*

Power line communications (PLCs) are to use existing electrical wires to transport data. Recently, new PLC technologies are available that allow high bit rates of up to 200 Mb/s. PLC can be used in several important applications: broadband Internet access, indoor wired local area networks, utility metering and control, real-time pricing, distributed energy generation, etc. [2].

From a standardization point of view, competing organizations have developed specifications, including HomePlug Powerline Alliance, Universal Powerline Association and HD-PLC. ITU-T adopted Recommendation G.hn/G.9960 as a standard for high-speed power line communications. In IEEE, P1901 is a working group developing PLC medium access control and physical layer specifications. National Institute of Standards and Technology (NIST) has included HomePlug, ITU-T G.hn and IEEE 1901 as "Additional Standards Identified by NIST Subject to Further Review" for the smart grid in the USA [1].

The primary advantage of PLC arises from the fact that it allows communication signals to travel on the same wires that carry electricity. However, since power line cables are often unshielded and thus become both a source and a victim of electromagnetic interference (EMI). Another issue is the price. A PLC module is usually more expensive than a wireless module, such as ZigBee, which will be introduced in the next subsection. In addition, wireless is also more practical in some applications, such as water/gas meters powered by batteries without power lines.

*B. Wireless Home (Local) Area Networks*

A leading standard for the wireless home network communications is ZigBee. The Zigbee Smart Energy standard builds on top of the ZigBee Home Automation (HAN) standard. HAN provides a framework to automatically control lighting, appliances, and other devices at home. ZigBee Smart Energy provides a framework to connect HAN devices with smart meters and other such devices. This will enable the energy utility to directly communicate with the end consumers of energy.

Wi-Fi is often used as a synonym for IEEE 802.11 wireless local area network (WLAN) technologies. Recently, Wi-Fi became a standard for laptops and subsequently phones due to its high data rate. When using it in utilities, however, Wi-Fi's power consumption is an issue that needs to be considered carefully.

The ZigBee Alliance and the Wi-Fi Alliance also consider collaborating on applications for energy management and networking. The initial goal will be to get Smart Energy 2.0, a standard promoted by ZigBee, to work on Wi-Fi.

*C. Wireless Wide Area Networks*

Public cell phone carriers have great interest in using wireless wide area networks to connect household smart meters directly with the utility's systems. A major advantage of this approach is the reduction of the costs (by not having to build a new network and by leveraging the expertise of the telecom world). However, since public wireless cellular networks are not specialized in machine-to-machine area, some requirements in utilities may not be met by cellular networks. Others argue that if the public wireless giants want to get into this business, they will do whatever it takes to meet the requirements to win these large-scale, multi-year utility contracts.

WiMAX is based on the IEEE 802.16 standard, enabling the delivery of wireless broadband communications. Unlike the now-popular wireless networking technologies using unlicensed spectrum (such as those used by Silver Spring and Trilliant), WiMAX uses licensed wireless spectrum, which is arguably both more secure and reliable. The primary disadvantage of using a licensed network is that it is more expensive. In addition, compared to cellular technologies, WiMAX has yet to be deployed at scale, which means some risks when applied to utilities.

*D. Interoperability of Different Communication Systems*

Without a framework of interoperable standards for communications, it would be very difficult to integrate renewable energies into the grid. Since the potential standards landscape is very large and complex, interoperable standards adoption is challenging. Many utilities and regulatory groups are collectively trying to address interoperability issues through workgroups such as the GridWise Architecture Council and Open Smart Grid (Subcommittee of the Utility Communications Architecture International Users Group) as well as through policy action from NIST. In June 2009, NIST announced an interoperability project via IEEE P2030, which seeks to define interoperability of energy technology and information technology operations with electric power systems and end-user applications and loads [3].

III. COMMUNICATION SYSTEMS FOR GRID INTEGRATION OF WIND FARMS

Wind energy has become an increasingly significant portion of the generation mix. Large scale wind farms are normally integrated into power *transmission* networks so that the generated electric power can be delivered to load centers in remote locations. Small scale wind farms can be integrated into power *distribution* networks to meet local demands. Because of high variability and intermittency, wind farm operations become a great challenge to power systems [4].

Communication systems are fundamental infrastructure that transmits measured information and control signals between wind farms and power systems. Well-designed communication systems can better explore the wind potentials and facilitate farm controls, helping shaving peak load and providing voltage support for power systems. Any deficiency in communication systems could compromise the system observability and controllability, which would negatively impact system security, reliability, and safety.

In this section, we present the communication systems used in a real renewable energy project, Bear Mountain Wind farm (BMW) in British Columbia, Canada.

A.  *Introduction to Bear Mountain Wind Farm* (*BMW*)

Located in the Peace River area of northern interior British Columbia, Canada, Bear Mountain Wind farm (BMW) consists of thirty four ENERCON wind turbine generators with a generation capacity of 102 MW. It is the first large-scale wind farm that has been integrated into British Columbia's transmission network [5]. In commercial operation since Dec. 2009, BMW is a typical project in a sense that it integrates large-scale wind resources into bulk power systems by adopting up-to-date technologies. The reliable and flexible operation of BMW, including active power coordination, reactive power control, wind farm protection and system protection, has been supported by the communication infrastructure specifically designed for this project.

Figure 1 shows a high level scheme of grid integration of BMW to bulk power system, where information flow and energy flow among BMW, transmission system, distribution system and generation system are summarized in an illustrative way. It can be seen that a modern power system is composed of high power equipment and communication networks. Energy flows through the power grid to meet customer demand, while information flows through the communication system to monitor the system status, control the dynamic energy flows presented in the grid, and transfer the information collected from an internet of smart devices for sensing and control across the power grid.

*B. Communication Systems for BMW Supervisory Control and Data Acquisition (SCADA)*

Inside BMW, the wind farm SCADA system is used for data acquisition, remote monitoring, open-loop and closed-loop control for both individual wind turbines and the whole farm. BMW SCADA also provides a platform for the customer, the manufacturer (ENERCON), and the utility (BC Hydro) to access the operating state and to analyze sampled event data. Moreover, authorized users can use the SCADA to modify parameters of wind energy converter (WEC) controllers and voltage control system (VCS), etc. This feature is of special importance because wind farm controllers need to be tuned to achieve optimized performances under varying power system conditions. The closed-loop control to regulate the voltage at Point of Interconnection (POI) is another desired SCADA feature, which coordinates wind turbine outputs and provides reactive power support for utility system. Figure 2 shows a high level design of the wind farm SCADA communication system and interface with external communication systems, which are briefly explained as follows.

- SCADA REMOTE: Used for remote monitoring of wind farm data. Authorized users may access SCADA database and modify controller parameters.

- PDI (Process Data Interface): Used for exchanging real time wind farm data with external communication systems.

- GDA (Grid Data Acquisition): Used to measure electrical variables at the point of interconnection.

- SCU (Substation Control Unit): Used for monitoring electrical states and for remote switching operation within the substation of wind farm.

- VCS (Voltage Control System): Used for controlling the dynamic voltage at the point of interconnection by utilizing reactive power capability of wind turbines online.

- METEO: Used for collecting meteorological data such as wind speed, wind direction, temperature, etc.

The status data and measured data transmitted by SCADA using standard formats such as OPC XML-DA, or IEC 60870-5-101(or 104), etc. BMW SCADA cyclically queries the operating and status data from wind turbines via the data bus. The SCADA calculates the average values over 10-minute, day, week, month and year periods, together with minimum and maximum values. Status data is updated up to four times per second. The internal data bus system normally uses fiber optical cables to guarantee the communication speed.

When the SCADA communication system is forced out of service or control signals to/from individual wind turbine/generators are interrupted, each affected generator output will automatically default to autonomous voltage control at 15% of its capacity and 1.0 power factor.

It should be emphasized here that the SCADA system response in the event of communication breakdowns has to be determined case by case.

*C. Communication System for Power System Protection & Control and Remedial Action Schemes*

The grid integration of BMW involves the construction of a new 138kV substation at the POI. The existing 138 kV line, 1L362 between Chetwynd (CWD) and Dawson Creek (DAW) has been looped in and out of the new station and has been split into two new lines: 1L358 and 1L362. At Chetwynd (CWD) and Dawson Creek (DAW) substations, the existing protection, control and metering devices for the old 1L362 line has been replaced with new protection and control equipment. At CWD substations, a remote terminal unit (RTU) using a 2400bps continuous SCADA channel will be installed. All protection information is transferred from SEL relays to the new RTU via SEL 2032 communication processor, and then sent to the utility control centre from this RTU through power line carriers.

To facilitate the new protection devices and to transfer control/telemetry/alarms data, new power line communication (PLC) systems are installed at POI substation, CWD, DAW and neighboring Peace Canyon substation.

Another important protection system implemented for BMW is the Remedial Action Scheme (RAS) [6]. Generally speaking, RAS is a specialized protection system which reacts to predictable power system contingencies by using pre-planned control actions. The purpose of RAS is to mitigate the unwanted consequences of the initiating condition or disturbance. Unlike the traditional protection system that only protects single equipment from local faults, RAS can protect multiple equipment located remotely from the initiating condition, and can improve reliability on a whole system level. Nowadays, power system reliability highly relies on the dependable performance of RAS, again, RAS functionality largely depends on the robustness of communication systems used. Communication system is designed to facilitate the RAS implementation as follows.

The communication channels provided for the RAS scheme consist of single mode fiber optic pairs on the ADSS line between POI and BMW. SEL 2800 series media converters are used for transmission of the Mirrored Bits information between the SEL 421 relays. Another fiber pair on the same ADSS cable is utilized for the BMW SCADA channel using the SEL 2800 series media converters. The BMW SCADA channel is cross connected to the ETL 600 Power Line Carrier Link in the direction from CWD to GMS substation where it cross connects to the Alcatel 3600 DACS and microwave radio in the direction of Williston which provides aggregate function for all the Northern utility SCADA channels. The aggregated traffic is then directed to the utility control center.

BMW data together with protection information and line telemetry data are transmitted to the system control center through ADSS fiber cable and power line carrier. Every 4 seconds, the data will be updated. Figure 3 illustrate part of BMW information visualized in the Energy Management System in BC Hydro's control center.

*D. Research Challenges*

- *Standardization of protocols*: As we can see from above description, there are many different communication devices for different purposes in a wind farm. A uniform communication platform for monitoring, control and operation of wind farms is needed such that the communication barriers arising from many proprietary protocols used by different manufacturers could be reduced to a minimum level [7]. Scalability and interoperability among communication equipment will minimize the maintenance effort and improve communication availability.

- *Implementation of synchronized phasor measurement*: A normally overlooked technology for wind farm applications is the synchronous phasor measurement. Wind energy is highly variable and intermittent, which requires high speed accurate power system dynamic control to counteract to the sudden changes in wind output. The correct real-time control relies on accurate estimation of system states which is otherwise impossible without synchronous phasor measurement units [8]. Developing a communication framework for synchrophasors and management of the synchrophasor data will be one of the major challenges to be resolved.

- *Application of wireless technologies*: So far the grid integration of wind energy mainly utilizes wired communications such as PLC, optical fiber, copper wires, etc. Using wireless communications for distributed monitoring, authorization, and control may significantly improve wind generation reliability and efficiency, and reduce the life cycle cost of wind power projects.

- *Make use of full capabilities of wind farm SCADA and wind turbine reactive capability*: Many advanced applications in wind farm SCADA, for example voltage control systems, are often unexplored by utilities because of communication barriers. Developing robust two-way communications can activate these useful functions so that wind energy efficiency, control speed, and support to grid can be greatly improved.

- *Enhance communication systems reliability*: Communication systems failure can result in limited operation of wind farms. The wind farm output may have to be reduced in order to guarantee power system reliability and safety. Therefore, high reliability of communication systems can increase the wind energy yield which is beneficial for not only wind farm owners, but also utilities and customers.

- *Islanding detection and operation through communication systems*: An electrical island forms when a portion of the power system becomes electrically isolated from the rest of the system, while it continues to be energized by wind farms or other distributed generators. An energized island may cause severe safety hazards and must be effectively detected before some control measures can be taken to fix the problem. The traditional method for islanding detection is to monitor the frequency and/or voltage drifts, which may not function properly or fast enough. Using wireless communications or PLC could be a fast and accurate means for island detection and control.

IV. Communication systems for Grid Integration of Photovoltaic Power Systems

A. *Introduction to Photovoltaic Power Systems* (*PPS*)

The first application of photovoltaic power was as a power source for space satellites. Today, the majority of photovoltaic modules are used for utility-interactive power generation. Grid-connected solar systems are typically classified as three categories: residential, commercial, and utility scales. Residential scale is the smallest type of installation and refers to all installations less than 10kW usually found on private properties. The commercial capacity ranges from 10kW to 100kW, which are commonly found on the roof of a commercial building. Utility scale is designed to the installations above 100kW, which are traditionally ground-based installations on fields (also known as solar farms or plants).

*B. Communication Systems for PPS Supervisory Control and Data Acquisition (SCADA)*

Data acquisition was originally found in the utility scale solar power systems for monitoring connection status, real and reactive power output, and voltage at the interconnection point. With the spread of the Internet, web-based tools have been developed to give photovoltaic system owners access to the current and historical data of their systems. Since then, the communication system becomes cost-effective and applicable to small-scale power systems. As shown in Figure 4, the monitoring interface gets electrical generation data from the inverters and transmits it to the server via the Internet. Some systems also provide sensors to collect data of ambient temperatures, solar irradiance, total generation, and usage data from the electrical panel. RS-485 is widely used as the protocol between the inverters and the data logger. Ethernet and Internet are typically the media for local and remote monitoring. A RS232 or USB interface is handy for the on-site debug, configuration, or monitoring for one-inverter systems.

These systems normally monitor the following variables: the solar array power production, inverter output, inverter status, AC grid conditions, weather station data, temperature of key components, and solar irradiance, etc. The owners or operators can follow the real-time details about the system operation. Some monitoring data are metering-quality and used for feed-in tariff.

*C. Communication Systems for PPS Advanced Applications: Fault Diagnosis*

A photovoltaic cell is a device that converts light energy into electrical energy. Photovoltaic cells are often encapsulated as a module. Most solar power installations are made up of several strings, each of which includes two or more photovoltaic modules. Strings are then interconnected to create an array with the desired current or power.

It is shown in [9] that non-optimal conditions, such as minor shading, can cause a major reduction in solar power output of the photovoltaic array. Non-optimal conditions are sometimes unavoidable and complicated as they are caused by many different conditions: such as partial shading, soiling, dust collection, cell damage, cell ageing, etc. As a result, it is very difficult to detect a non-optimal condition by reading the power output or sensing the output terminal of a photovoltaic array.

A recent development is to integrate communication systems into the photovoltaic panel, to sense the voltage, current and temperature of each module, and send the information data to the monitoring interface. The solar power monitoring can be classified as three categories: system-level, string-level, and module-level. Figure 5 shows the three-level monitoring based on wireless communication systems. The system will monitor the status of solar modules, solar strings, and solar inverters based on the IEEE 802.15.4-2003 ZigBee standard. Either star or mesh topology can be used.

With this wireless monitoring capability, each solar module status is visible. In practical systems, this is very useful because most solar panels are installed in the areas that are not readily accessible. Otherwise, the troubleshooting of solar modules is very difficult.

*D. Research Challenges*

- *Power consumption of the end device*: A major challenge of photovoltaic power systems is to reduce the power dissipation losses. The power losses may be caused by non-optimal operational conditions, or by communication systems (e.g., wireless transmission and receiving). Developing an energy efficient communication network for solar power integration will significantly reduce the life cycle cost of solar projects.

- *Reliability, coverage, and flexibility*: A reliable communication system is the foundation for effective photovoltaic control. Although wireless communications provide great flexibility to the photovoltaic power systems control, the communication may be unreliable due to interference, shadowing, fading, etc. Moreover, the coverage of a wireless network changes dynamically in practice. These uncertainties in wireless communications can cause severe problems in photovoltaic power systems. The tradeoff among power consumption, reliability, coverage, throughput, latency, etc. merit further investigation for the wireless communications used in photovoltaic power systems.

- *Addressing and localization*: A solar power system consists of many photovoltaic panels, which are vulnerable to failures and have high maintenance demands. How to identify the failed panel quickly is an interesting research topic. Similar problems exist in wireless sensor networks, where a variety of addressing and localization algorithms have been designed. Adopting those addressing and localization algorithms for the communication systems used in photovoltaic power systems could provide possible solutions.

- *Islanding detection*: Photovoltaic power systems face the same challenge of island detection as wind power systems do. Using wireless communications or PLC could be a fast and accurate means for island detection and control in photovoltaic power systems.

## V. CONCLUSIONS

Two-way communications are the fundamental infrastructure that enables the accommodation of distributed renewable energy generation. In this paper, we reviewed several communication technologies available for the grid integration of renewable energy resources. Since a hybrid mix of technologies will be used in the future, interoperable standards are very important. We introduced the Bear Mountain Wind Farm project, particularly its communication systems for supervisory control and data acquisition, as well as power protection and control. We also introduced the communication systems used in photovoltaic power systems. Distinct characteristics in integration of renewable energy resources pose new challenges to the communication systems, which merit further research.

[8] IEC 61400-12-1, Wind turbines – Part 12-1: Power Performance Measurements of Electricity Producing Wind Turbines.

[9] W. Xiao, N. Ozog, and W. G. Dunford, "Topology study of photovoltaic interface for maximum power point tracking," *IEEE Trans. Industrial Electronics*, vol. 54, no. 3, pp. 1696-1704, June 2007.

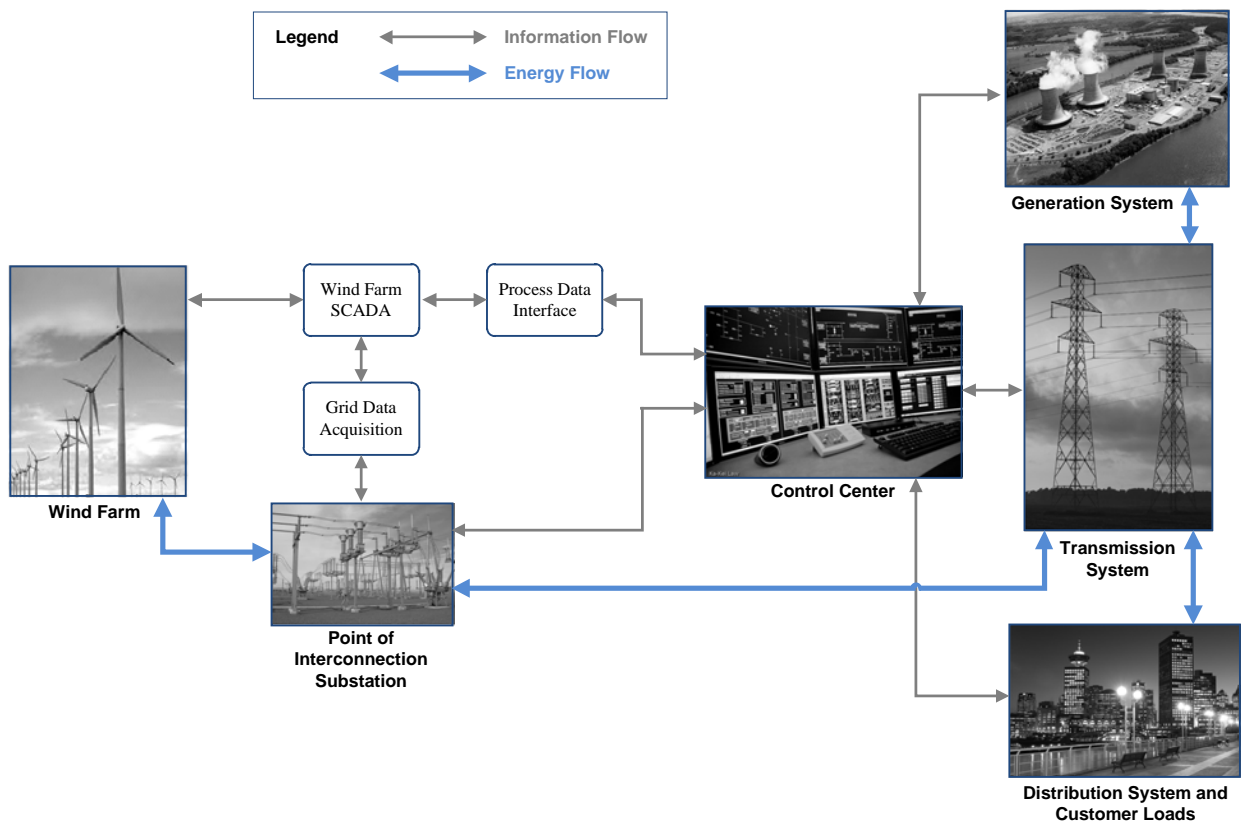

Figure 1. Grid integration of a wind farm. (Photo courtesy of BC Hydro)

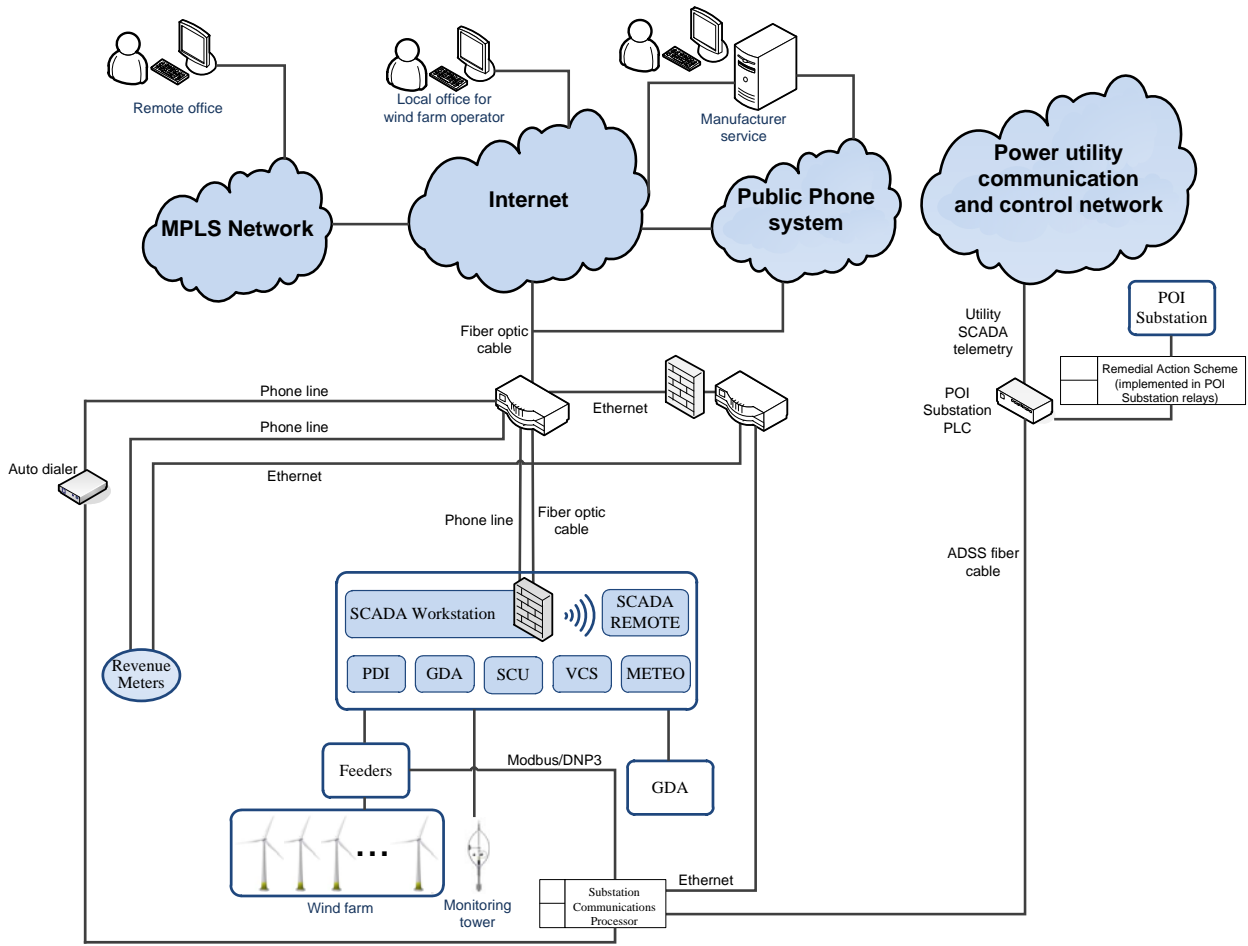

Figure 2. A typical SCADA communication system in a wind farm.

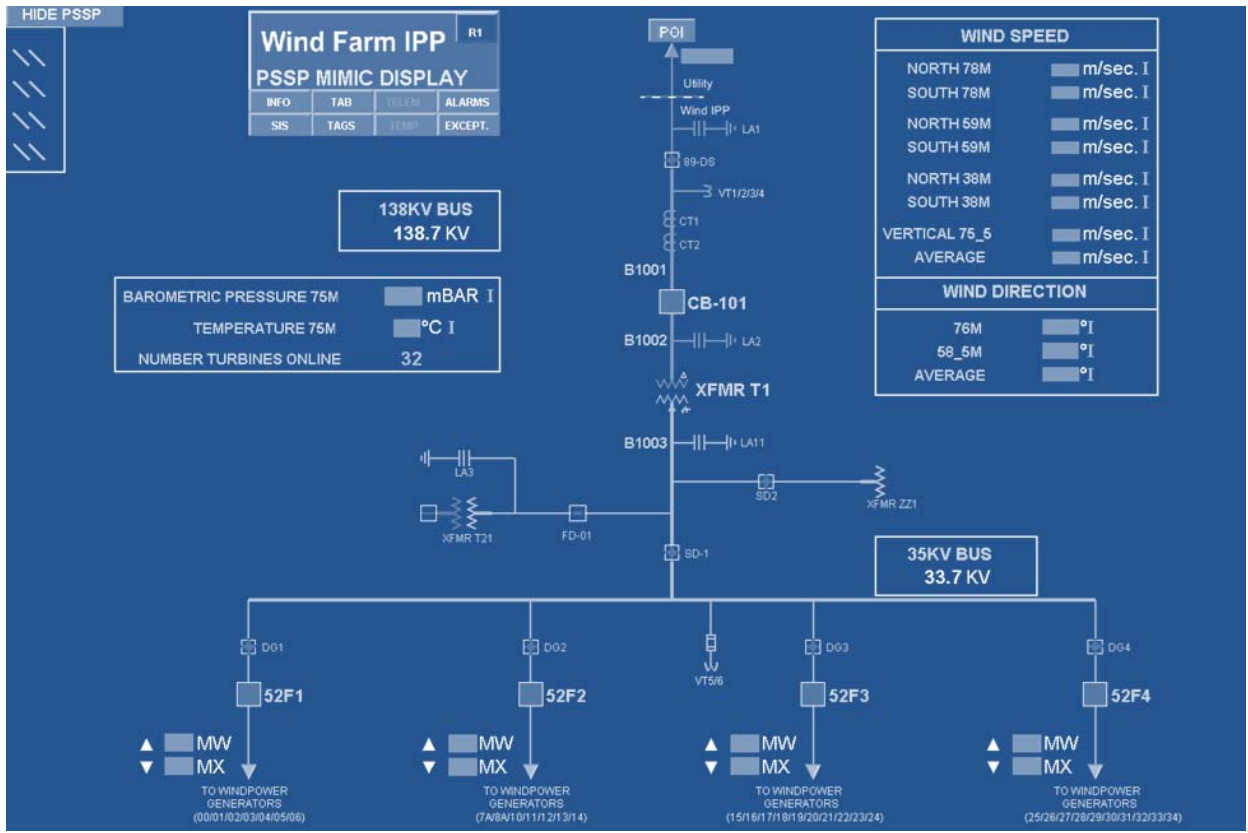

Figure 3. One page of wind farm information on utility company SCADA display (Data masked).

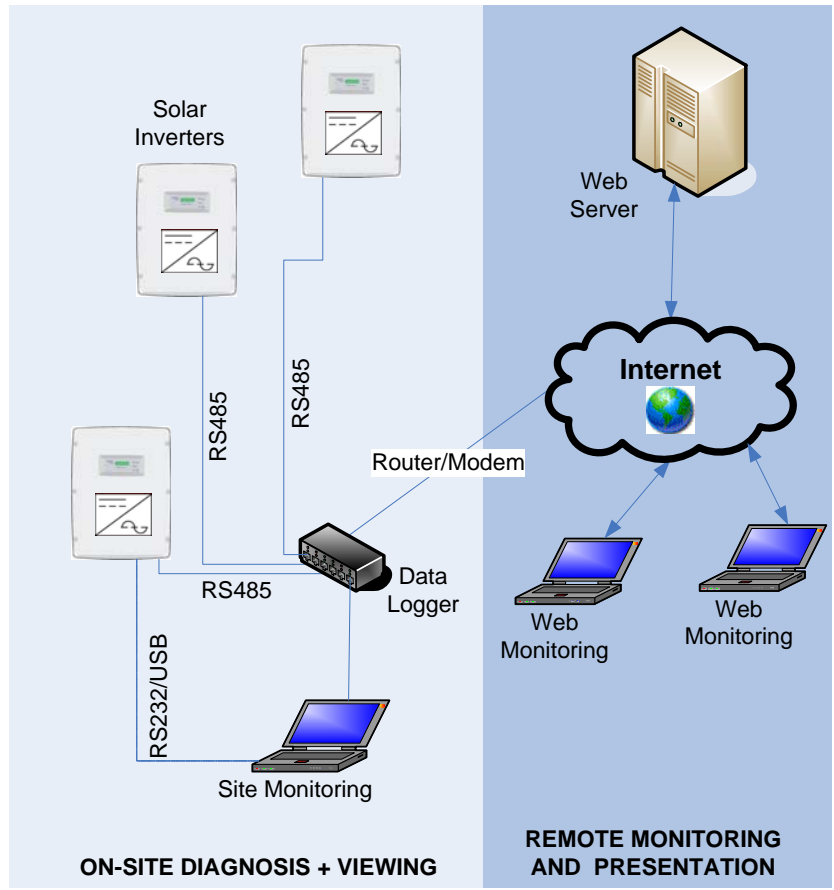

Figure 4. A typical monitoring application of grid-tied photovoltaic power systems.

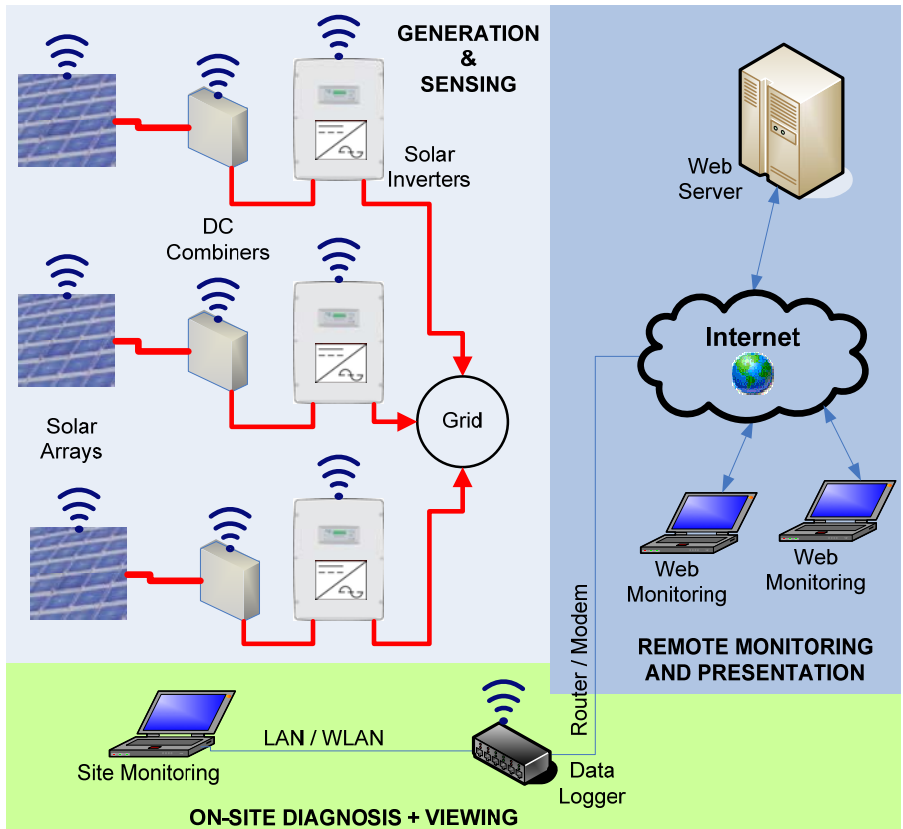

Figure 5. Three-level monitoring of photovoltaic power systems based on wireless communication technologies.

21